\documentclass[aps, prd, onecolumn, tightenlines, notitlepage, superscriptaddress, nofootinbib, preprintnumbers, floatfix,showkeys,11pt,altaffilletter]{revtex4-2}
\usepackage{upgreek}

\usepackage[official]{eurosym}
\usepackage[normalem]{ulem}
\usepackage{amstext}
\usepackage{graphicx}
\graphicspath{{figuresv2/}}
\usepackage{url}
\usepackage{color}
\usepackage{ulem}
\usepackage[version=4]{mhchem}
\usepackage[utf8]{inputenc}
\usepackage{fontawesome}
\usepackage{yfonts}
\usepackage{mathrsfs}

\pdfoutput=1
\usepackage{textcomp}
\usepackage{comment}
\usepackage{yfonts}
\usepackage{epsfig,amsfonts,mathrsfs,graphicx,color,slashed,multirow}
\usepackage{latexsym,graphicx,slashed,color,enumerate,url,cancel,gensymb}
\usepackage{textcomp}

\usepackage[x11names]{xcolor}
\usepackage[colorlinks,pdfstartview=FitV,breaklinks=true]{hyperref}
\usepackage{booktabs}
\usepackage{adjustbox}
\usepackage{textgreek} 
\usepackage{caption, subcaption} 
\usepackage{booktabs} 
\usepackage{float}

\usepackage{lmodern}
\usepackage{ae,aecompl}
\usepackage{appendix}
\usepackage{orcidlink}
\usepackage{nccmath} 
\usepackage{textcomp}
\allowdisplaybreaks


\definecolor{vdrgreen}{rgb}{0.0, 0.6, 0.0}

\definecolor{magenta(dye)}{rgb}{0.79, 0.08, 0.48}
\definecolor{byzantium}{rgb}{0.44, 0.16, 0.39}

\DeclareMathOperator{\diag}{diag}

\makeatletter
    \newcommand{\colorboxed}[3][white]{\fcolorbox{#2}{#1}{\m@th$\displaystyle#3$}}
\makeatother

\AtBeginDocument{\hypersetup{citecolor=magenta(dye),linkcolor=magenta(dye),urlcolor=magenta(dye)}}
\usepackage{appendix}

\begin{document}

\title{Probing damping effects in neutrino oscillations with 207 days of JUNO data}

\author{Martina Beccaria}
\email{martina.beccaria@gssi.it}
\affiliation{Gran Sasso Science Institute, Viale F. Crispi 7, L’Aquila, 67100, Italy}
\affiliation{Istituto Nazionale di Fisica Nucleare (INFN), Laboratori Nazionali del Gran Sasso, 67100 Assergi, L’Aquila (AQ), Italy
}
\author{Christoph A. Ternes}
\email{christoph.ternes@lngs.infn.it}
\affiliation{Gran Sasso Science Institute, Viale F. Crispi 7, L’Aquila, 67100, Italy}
\affiliation{Istituto Nazionale di Fisica Nucleare (INFN), Laboratori Nazionali del Gran Sasso, 67100 Assergi, L’Aquila (AQ), Italy
}

\begin{abstract}
We consider different scenarios which lead to a damping of the neutrino oscillation probability and investigate their impact on the JUNO results. First, we study decoherence effects due to wave packet separation. In addition, we consider an open quantum system framework and adopt a phenomenological approach which allows us to parameterize the energy dependence of the decoherence effects more freely. Finally, we study the effect of invisible neutrino decay. In all cases with {207 days of} data, JUNO can already place competitive bounds on the parameter space of the scenarios under consideration, while also maintaining a relatively robust measurement of the standard neutrino oscillation parameters.
\end{abstract}
\maketitle

\section{Introduction}

Recently, the JUNO collaboration published their first result on neutrino oscillation measurements~\cite{JUNO:2025gmd}. With only 59 days of data they obtained the, at that moment, most precise measurements of $\sin^2\theta_{12}$ and $\Delta m_{21}^2$, surpassing the precision obtained in global fits to neutrino oscillation data~\cite{deSalas:2020pgw,Esteban:2024eli,Capozzi:2025wyn}.
This release inspired many phenomenological applications, covering implications for global analyses~\cite{Capozzi:2025ovi,Esteban:2026phq,Goswami:2025wla}, searches for physics beyond the standard model (BSM) in neutrino oscillations~\cite{Chattopadhyay:2025ccy,Araya-Santander:2025jfd,Choubey:2026jiq,Alves:2026ydc,Flores:2026vbx,Gonzalez-Alonso:2026sgl}, tests of unitarity~\cite{Xing:2025bdm,Huang:2025znh}, and also implications for flavor models~\cite{Chen:2025ruj,Calibbi:2025ded,Zhang:2025jnn,Ge:2025csr,Jiang:2025hvq,He:2025idv,Petcov:2025aci,Ding:2025dzc,Ding:2025dqd,Borah:2025vtn,Nanda:2025fvw,Shang:2026qkh,Dutta:2026dzh,Kumar:2026qee,Priya:2026tpk}. { Since then, also the second results have been published: in this case, with 207 days of data, the Collaboration was able to improve even further the precision of the solar parameters $\sin^2\theta_{12}$ and $\Delta m_{21}^2$ while also obtaining for the first time an indication on the neutrino mass ordering~\cite{JUNO:2026tap} (when using some external constraints)}.

Within the standard three-neutrino paradigm, the neutrino oscillation parameters are quite well measured, with the only remaining open issues being the octant of $\theta_{23}$, the value of the Dirac CP phase $\delta$ and the neutrino mass ordering (if $\Delta m_{31}^2>0$ or $\Delta m_{31}^2<0$). 
However, many models of physics beyond the standard model predict deviations from the standard oscillation picture. Given the precision of current oscillation data and the overall consistency among experiments, any BSM effects are expected to appear only as sub-leading corrections to the standard oscillation probabilities.

In this article, we extend the searches for new physics that can be addressed with JUNO data. There are several scenarios which lead to a damping of the neutrino oscillation probability. The JUNO collaboration previously investigated damping signatures at the sensitivity-study level~\cite{JUNO:2021ydg}. 
Here, we revisit some of these scenarios using the {latest} oscillation data released by JUNO. We are going to consider coherence loss caused by wave packet separation, neutrino decoherence due to an interaction with an unknown environment in the open quantum system framework, and the invisible decay of neutrinos. These scenarios lead to a similar phenomenology, but with different energy dependencies in the damping parameters. 
Neutrino experiments have already been proven to be a powerful tool to test these scenarios, see, for example, Refs.~\cite{Giunti:1991sx,Kiers:1997pe,Beuthe:2002ej,Blennow:2005yk,DayaBay:2016ouy,Akhmedov:2019iyt,deGouvea:2020hfl,Cheng:2020jje,deGouvea:2021uvg,deGouvea:2024syg} for the wave packet treatment, Refs.~\cite{Gago:2000qc,Gago:2002na,Benatti:2000ph,Benatti:2001fa,Lisi:2000zt,Morgan:2004vv,Anchordoqui:2005gj,Fogli:2007tx,Oliveira:2010zzd,Oliveira:2013nua,Bakhti:2015dca,Oliveira:2016asf,BalieiroGomes:2016ykp,BalieiroGomes:2018gtd,Coelho:2017byq,Coelho:2017zes,Carpio:2017nui,Carpio:2018gum,Coloma:2018idr,Carrasco:2018sca,Buoninfante:2020iyr,Gomes:2020muc,Ohlsson:2020gxx,Stuttard:2020qfv,Stuttard:2021uyw,deHolanda:2019tuf,ICECUBE:2023gdv,DeRomeri:2023dht,DEsposito:2023psn,Barenboim:2024wdn,ESSnuSB:2024yji,KM3NeT:2024jji,Ternes:2025mys,Acharya:2025rnw,Esteban:2026vzh} for quantum decoherence, and Refs.~\cite{Frieman:1987as, Ivanez-Ballesteros:2023lqa,Pagliaroli:2015rca,Denton:2018aml,Gonzalez-Garcia:2008mgl,Choubey:2018cfz,Lindner:2001fx,Abrahao:2015rba,Ghoshal:2020hyo,Chattopadhyay:2021eba,Chattopadhyay:2022ftv,Banerjee:2023sxj,Gomes:2014yua,Choubey:2017dyu,Ghoshal:2020hyo,Chakraborty:2020cfu,Abrahao:2015rba,Choubey:2020dhw,Choubey:2017eyg,deSalas:2018kri,KM3NeT:2023ncz,Ternes:2024qui,Martinez-Mirave:2024hfd} for invisible neutrino decay. 

Our paper is structured as follows: In Sec.~\ref{sec:analysis} we describe the data analysis procedure, in Sec.~\ref{sec:damping} we present the damping signatures that we will consider in a comparative way, in Sec.~\ref{sec:res} we discuss our results and, finally, in Sec.~\ref{sec:conc} we summarize and draw our conclusions.

\section{Analysis of JUNO data}
\label{sec:analysis}

JUNO, the Jiangmen Underground Neutrino Observatory, is a new reactor neutrino experiment. It is a 20-kton liquid scintillator detector located at approximately 52.5~km (see Ref.~\cite{JUNO:2021vlw} for the exact distances, which are also used in our simulations) from the Yangjiang (six reactors with a thermal power of 2.9~GW$_{\text{th}}$ each) and Taishan (two reactors with a thermal power of 4.6~GW$_{\text{th}}$ each) nuclear power plants. 
JUNO has collected data for {207.2 live days and observed already more than 8200 inverse beta decay events, after which they reported the measurements $\sin^2\theta_{12} = 0.3036\pm 0.0064$ and $\Delta m^2_{21} = (7.388\pm 0.078) \times 10^{-5}$ eV$^2$~\cite{JUNO:2026tap}}. 

In our analysis, we use the GLoBES package~\cite{Huber:2004ka,Huber:2007ji} for predicting the event spectrum at JUNO. {The calculation follows closely the description of Ref.~\cite{Forero:2021lax}.} The experimental details {(such as energy resolution, background components, reactor fluxes and cross-sections)} are taken from Refs.~\cite{JUNO:2021vlw,JUNO:2025gmd, JUNO:2026tap}. {Some important features will be briefly summarized in the following.
When calculating the event-numbers, we use an energy resolution of 3.5\%, as shown by the Collaboration in their calibration campaign. 
In addition to the nuclear power plants of Yangjiang and Taishan we included two effective reactors to take into account the Daya Bay ($L =215$~km, $P = 17.4$~GW$_\textrm{th}$) and the Fangchenggang ($L =412$~km, $P = 12.1$~GW$_\textrm{th}$) sites~\cite{Esteban:2026phq}, both at distances closer than 450~km.
Our statistical analysis accounts for an uncertainty on the normalization of every reactor of 1.2\% as shown in Table III of Ref~\cite{JUNO:2026tap}. Other uncertainties are taken from Ref~\cite{JUNO:2026tap} and Table II therein, and consist of 1.8\% for the overall oscillated flux normalization, 26\% for the $^9$Li/$^8$He background, 100\% for geoneutrinos, 10\% for other reactors not included in the signal, 71\% for $^{214}$Bi/$^{214}$Po background, 50\% for the other backgrounds.
We confront the resulting prediction on the oscillated energy spectrum} with the data collected at JUNO using the $\chi^2$ function 
\begin{equation}
\label{eq:chi2}
 \chi^2(\vec{p})=\min_{\vec{\alpha}}\left\{
 2\sum_i \left[ N_{\text{exp},i}(\vec{p},\vec{\alpha})- N_{\text{dat},i} +
 N_{\text{dat},i} \log \left(\frac{N_{\text{dat},i}}{N_{\text{exp},i}(\vec{p},\vec{\alpha})}\right)\right] 
 + \sum_i \left(\frac{\alpha_i}{\sigma_i}\right)^2\right\} + \chi^2_{\text{pen.}}\,,
\end{equation}
where $N_{\text{dat},i}$ and $N_{\text{exp},i}$ are the data and the prediction in energy bin $i$. The predicted number of events $N_{\text{exp},i}$ depends on the neutrino oscillation parameters, $\vec{p}$, while the vector $\vec{\alpha}$ contains the systematic uncertainties {discussed above}, both for the signal prediction and background contamination, which are penalized with $\sigma_i$ in the second term. 
{The last term contains a penalty. Since the measurements of standard oscillation parameters of certain data sets are not affected by the scenarios discussed in this paper, these may be included as external penalties in our analyses. The exact penalty will depend on the scenario under consideration and will be specified in the dedicated sections below.  
Our discussion is focused on the solar sector. However, the effects due to $\sin^2\theta_{13}$ and $\Delta m_{31}^2$ must be taken into account. Therefore, we minimize over these parameters, depending on the scenario either freely or using external constraints.}

\section{Damping signatures}
\label{sec:damping}
The $\overline{\nu}_e$ survival probability\footnote{Note that in our numerical calculations we also account for the matter potential.} relevant for JUNO can be written as 

\begin{equation}
P_{ee}
={}
|U_{e1}|^4 + |U_{e2}|^4 + |U_{e3}|^4 
+
2|U_{e1}|^2|U_{e2}|^2
\cos2\Delta_{21}\,
+
2|U_{e1}|^2|U_{e3}|^2
\cos2\Delta_{31}\,
+
2|U_{e2}|^2|U_{e3}|^2
\cos2\Delta_{32}\,,
\label{eq:osc_SM}
\end{equation}
where $\Delta_{ij} = \frac{\Delta m_{ij}^2 L}{4E}$, with $L$ and $E$ being the propagation baseline and neutrino energy, respectively. The elements of the mixing matrix are given by $|U_{e1}|^2 = c_{12}^2\,c_{13}^2$, $|U_{e2}|^2 = s_{12}^2\,c_{13}^2$ and $|U_{e3}|^2 = s_{13}^2$, where $s_{ij} = \sin\theta_{ij}$, $c_{ij} = \cos\theta_{ij}$.
If we denote with $D_{i(j)}(E,L)$ some generic damping terms {(which are functions of energy, baseline, and also of the fundamental parameters of the models considered in this work)}, the scenarios that we are going to investigate in our work will always be of the form

\begin{equation}
\begin{aligned}
P_{ee}^{\textrm{damp.}}
&=
|U_{e1}|^4 e^{-D_1(E,L)}
+
|U_{e2}|^4 e^{-D_2(E,L)}
+
|U_{e3}|^4 e^{-D_3(E,L)}
\\
&+
2|U_{e1}|^2|U_{e2}|^2
\cos2\Delta_{21}\,
e^{-D_{21}(E,L)}
\\
&+
2|U_{e1}|^2|U_{e3}|^2
\cos2\Delta_{31}\,
e^{-D_{31}(E,L)}
\\
&+
2|U_{e2}|^2|U_{e3}|^2
\cos2\Delta_{32}\,
e^{-D_{32}(E,L)} .
\end{aligned}
\label{eq:osc_damp}
\end{equation}
As can be seen, it maintains the same structure as Eq.~\eqref{eq:osc_SM}, but each term is now multiplied with a damping term. In the following subsections we will briefly introduce the scenarios which are going to be studied in this paper.
The effect of the different scenarios on the oscillation probability is shown in Fig.~\ref{fig:probs} and will be discussed in the following subsections. In all panels we use the parameters from Ref.~\cite{deSalas:2020pgw} for the standard neutrino oscillation parameters.

\subsection{Neutrino wave packets}
\label{sec:wave}

The first scenario that we consider is neutrino decoherence due to wave packet separation. Neutrinos can lose coherence due to the separation of mass states in propagation (see e.g. Refs.~\cite{Giunti:1991sx,Giunti:2003ax,Beuthe:2001rc,Beuthe:2002ej,Akhmedov:2019iyt}). Once the wave packets associated to the different mass states stop overlapping, coherence is lost and oscillations become suppressed. In this case, the oscillation probability is given by Eq.~\eqref{eq:osc_damp} with

\begin{equation}
    D_i(E,L) = 0\quad(i=1,2,3)\,,\quad D_{ij}(E,L) = \left(\frac{ \Delta m_{ij}^2 L}{4\sqrt{2}E^2\sigma}\right)^2\quad(i\neq j)\,,
    \label{eq:D_wavepacket}
\end{equation}
where $\sigma$ is the width of the wave packet. Note that theoretical expectations of the neutrino wave packet width are still the subject of debate. However, they seem to point towards larger values than those accessible by current experiments of all types~\cite{Akhmedov:2022bjs,Jones:2022hme,Krueger:2023skk}. Therefore, if we measured a finite value of $\sigma$ it would definitely hint towards either new physics or to an incomplete understanding of the coherence of neutrino sources. 

\begin{figure}[t!]
    \centering
    \includegraphics[width=0.49 \textwidth]{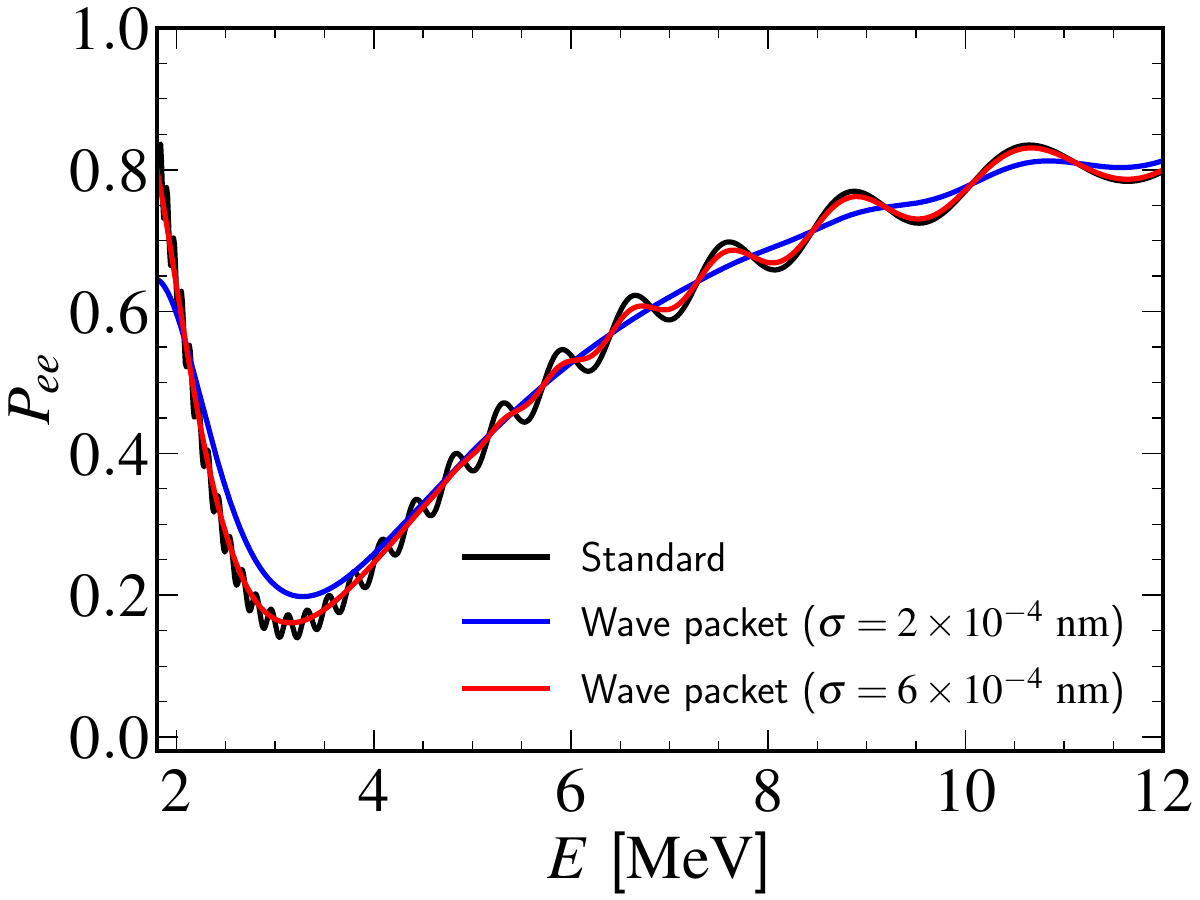}
    \includegraphics[width=0.49 \textwidth]{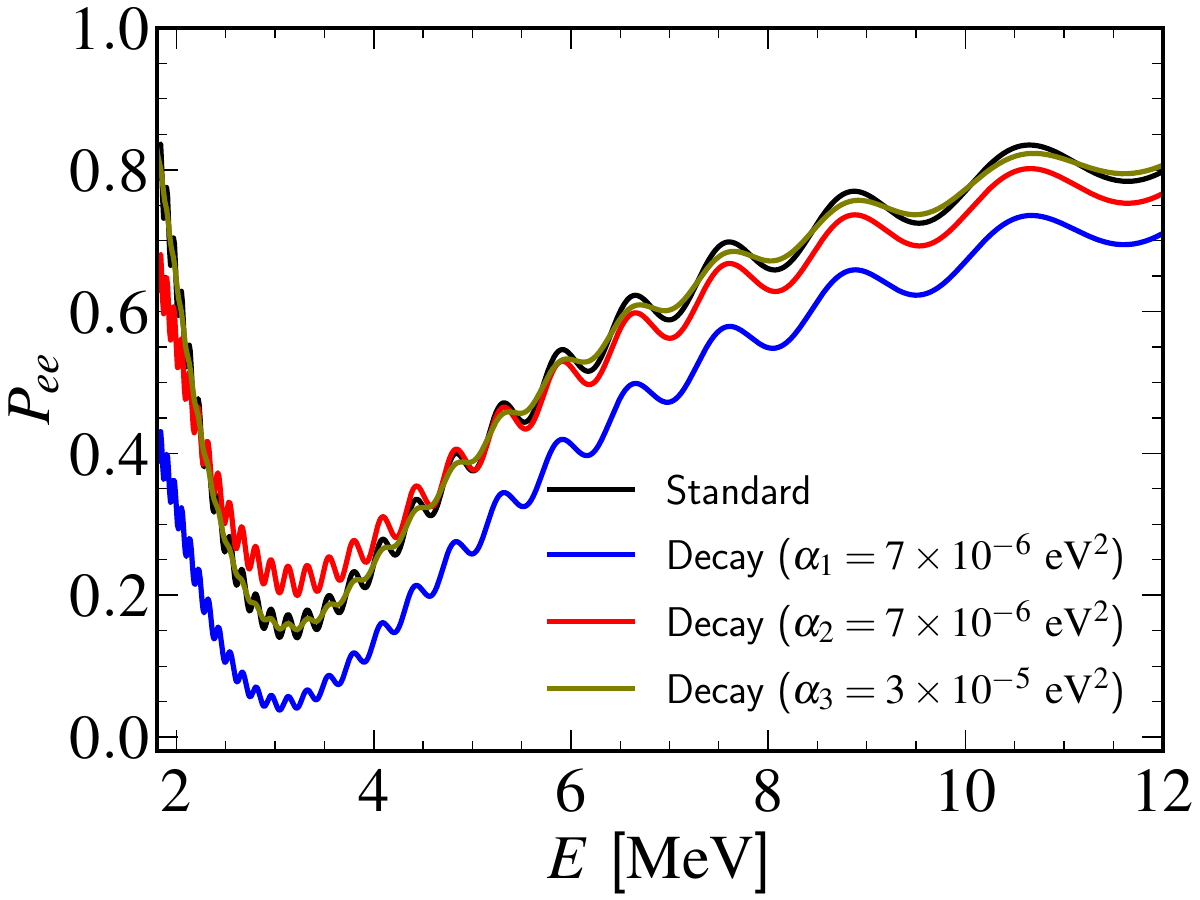} \\
    \includegraphics[width=0.49 \textwidth]{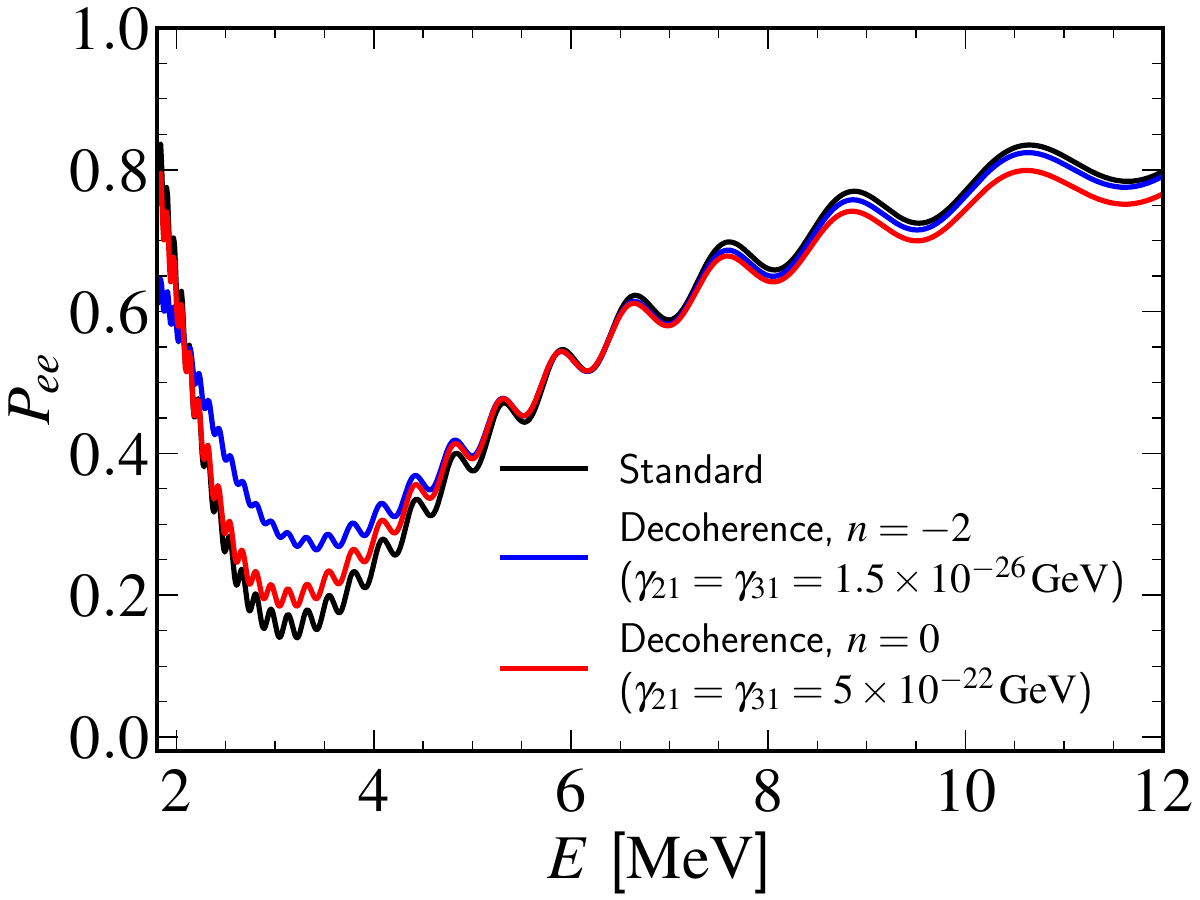}
    \includegraphics[width=0.49 \textwidth]{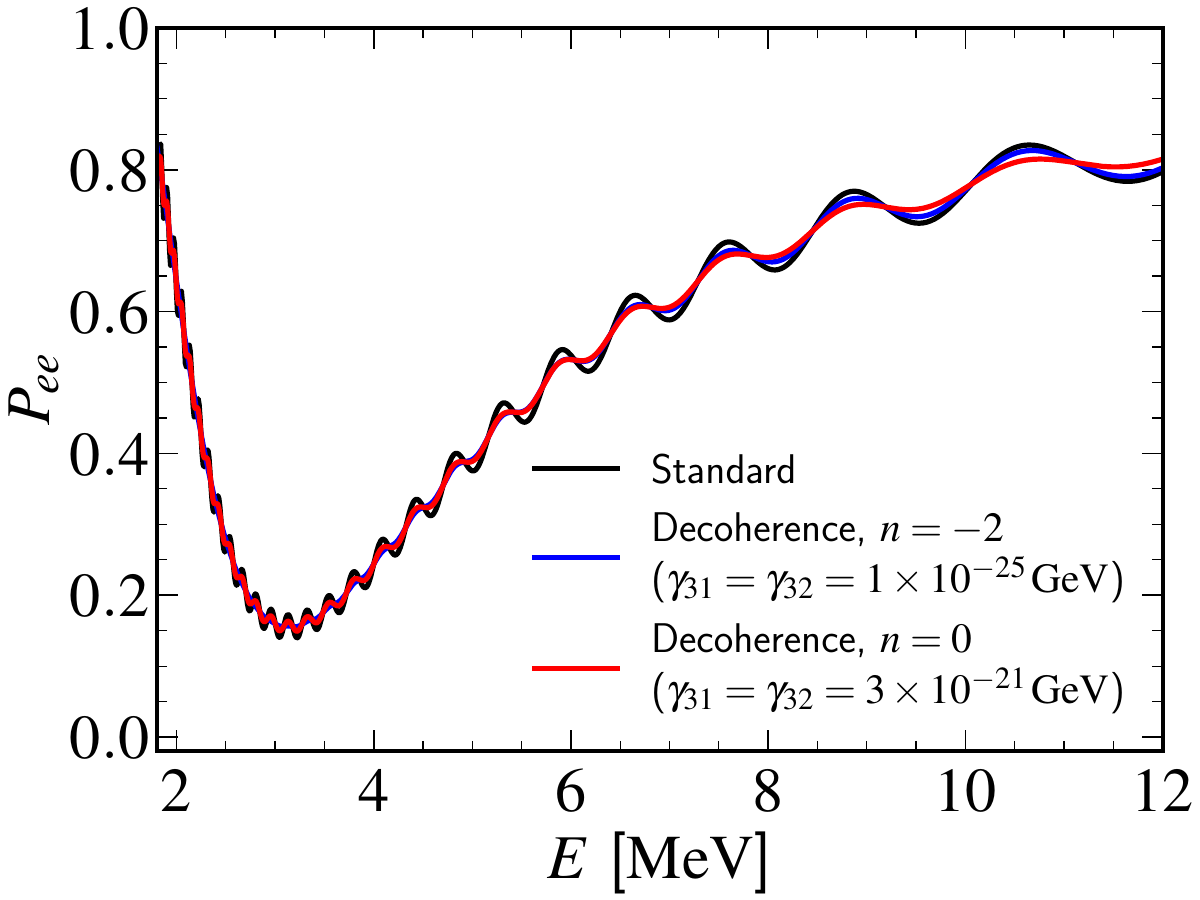}
    \caption{The effect of wave packet separation (upper left panel), neutrino invisible decay (upper right panel) and environmental decoherence (lower panels, left for $\gamma_{21}=\gamma_{31}, \,\, \gamma_{32}=0$ and right for $\gamma_{31}=\gamma_{32}, \,\, \gamma_{21}=0$) on the neutrino oscillation probability at JUNO (fixing $L=52.5$~km).}
    \label{fig:probs}
\end{figure}

In the upper left panel of Fig.~\ref{fig:probs} we show how wave packet separation washes out {first the fast oscillation pattern, before beginning to affect also the slow oscillation. We show in black the standard probability and in blue and red the ones obtained for a wave packet width of $\sigma=2 \times10^{-4}$~nm and $\sigma=6 \times10^{-4}$~nm, respectively. As can be seen at $\sigma=6 \times10^{-4}$~nm at large neutrino energies the fast oscillations are still present, while at low energies they are washed out. At $\sigma=2 \times10^{-4}$~nm the fast oscillations are fully gone and wave packet separation begins to affect also the slow oscillation.}

\subsection{Neutrino environmental decoherence}
\label{sec:lindblad}

Neutrinos can also lose coherence if they interact with some (unknown) environment. In this scenario, the flavor evolution is described through the Lindblad Master-equation~\cite{Lindblad:1975ef,Gorini:1975nb}

\begin{equation}
	\frac{\partial\rho_\nu(t)}{\partial t}=-i[H,\rho_\nu(t)] + \mathcal{D}[\rho_\nu(t)]\,,
    \label{eq:rho_time_dep}
\end{equation}
where $\rho_\nu(t)$ is a Hermitian density matrix that describes the neutrino states in the mass basis and $H$ is the Hamiltonian of the neutrino subsystem. The dissipative term $\mathcal{D}[\rho_\nu(t)]$ encodes the decoherence effects. Under certain assumptions, see e.g. Refs.~\cite{BalieiroGomes:2016ykp,Oliveira:2016asf,BalieiroGomes:2018gtd,Lisi:2000zt,Fogli:2007tx,Coloma:2018idr,Guzzo:2014jbp,Carrasco:2018sca,Gomes:2020muc,Stuttard:2020qfv,DeRomeri:2023dht}, this formalism leads to Eq.~\eqref{eq:osc_damp} with

\begin{equation}
    D_i(E,L) = 0\quad(i=1,2,3)\,,\quad D_{ij}(E,L) = \gamma _{ij}L\left(\frac{E}{E_0}\right)^n\quad(i\neq j)\,,
    \label{eq:D_lindblad}
\end{equation}
where for easy comparison with the literature we fix $E_0 = 1$~GeV. Since the energies of reactor neutrinos are much smaller than this pivot energy, we consider only the cases with $n=-2,-1,0$. 
The case of $n=-4$ is not directly relatable to the decoherence-loss due to wave packet separation. While the energy dependence would coincide, note that Eq.~\eqref{eq:D_lindblad} depends on $L$, while for the wave packet separation the damping term in Eq.~\eqref{eq:D_wavepacket} depends on $L^2$.

In the lower left panel of Fig.~\ref{fig:probs} we compare the oscillation probabilities for $n=0$ (red) and $n=-2$ (blue) for $\gamma:=\gamma_{21}=\gamma_{31}$ and $\gamma_{32} = 0$. In contrast to the case of wave packet separation, the $\Delta m_{32}^2$ oscillation is not affected and the fast oscillations survive\footnote{Note that fixing $\gamma:=\gamma_{21}=\gamma_{32}$ and $\gamma_{31} = 0$ would lead to essentially the same pattern, and hence, we do not consider this case in the discussions below.}.
For $n=0$ the decoherence terms only change the amplitude of the slow oscillation term, while in the case of $n=-2$ we also see a change in frequency. We can therefore expect correlations only with $\sin^2\theta_{12}$ in the energy-independent case, while also with $\Delta m_{21}^2$ in the energy-dependent scenarios.       

{In the lower right panel of Fig.~\ref{fig:probs} we compare the oscillation probabilities for $n=0$ (red) and $n=-2$ (blue) for $\gamma:=\gamma_{31}=\gamma_{32}, \,\, \gamma_{21}=0$. 
Here the slow oscillation is not affected while the $\Delta m_{31}^2$ and $\Delta m_{32}^2$ oscillation patterns might be washed out due to the presence of the decoherence term. 
We can therefore expect limited correlations with the solar parameters in the three scenarios considered.}

\subsection{Neutrino invisible decay}
\label{sec:decay}
Finally, we consider invisible neutrino decay. We can assume that active neutrinos are subject to both standard mixing and invisible decays~\cite{Lindner:2001fx,Abrahao:2015rba,Choubey:2018cfz,Ternes:2024qui}. In this hypothesis, the evolution process is described by the Hamiltonian given by
\begin{equation}
 H = \frac{1}{2E} \left[H_0 + H_M + H_D\right],
 \label{Ham_decay}
\end{equation}
where the first two terms correspond to standard vacuum oscillations and matter effects, while the last term represents the neutrino decay part
\begin{equation}
 H_D = U\diag(-i\alpha_1,-i\alpha_2, -i\alpha_3)
U^\dagger .
\end{equation}
It can be shown that this leads again to Eq.~\eqref{eq:osc_damp}, this time with

\begin{equation}
    D_i(E,L) = \alpha_i\frac{L}{E}\quad(i=1,2,3)\,,\quad D_{ij}(E,L) = \frac{D_i(E,L)+D_j(E,L)}{2}\quad(i\neq j)\,,
    \label{eq:D_decay}
\end{equation}
with $\alpha_i = \frac{m_i}{\tau_i}$, where $m_i$ and $\tau_i$ are the mass and lifetime of $\nu_i$. The invisible neutrino decay scenario is similar to Eq.~\eqref{eq:D_lindblad} with $n=-1$. Note, however, that due to the presence of $D_i(E,L)$ the evolution is not unitary (unlike in the other two cases) and the neutrino oscillation probability is not conserved anymore.

In the upper right panel of Fig.~\ref{fig:probs} we show the oscillation probability in presence of $\alpha_1$ (blue), $\alpha_2$ (red) {and $\alpha_3$ (olive)}. We see that even though invisible decay induces also spectral distortions, the dominant effect is an overall suppression of the survival probability {when considering a decay of $\nu_1$ or $\nu_2$, while the effect of a $\nu_3$ decay is very limited and affects only the fast oscillations}.

\section{Results}
\label{sec:res}
In the following subsections we discuss the results obtained for the different scenarios considered in this paper. We discuss the bounds on the damping parameters and also the robustness of the measurement of the solar neutrino oscillation parameters.

\subsection{Neutrino wave packets}
\label{sec:res_wave}

\begin{figure}[t!]
    \centering
    \includegraphics[width=1 \textwidth]{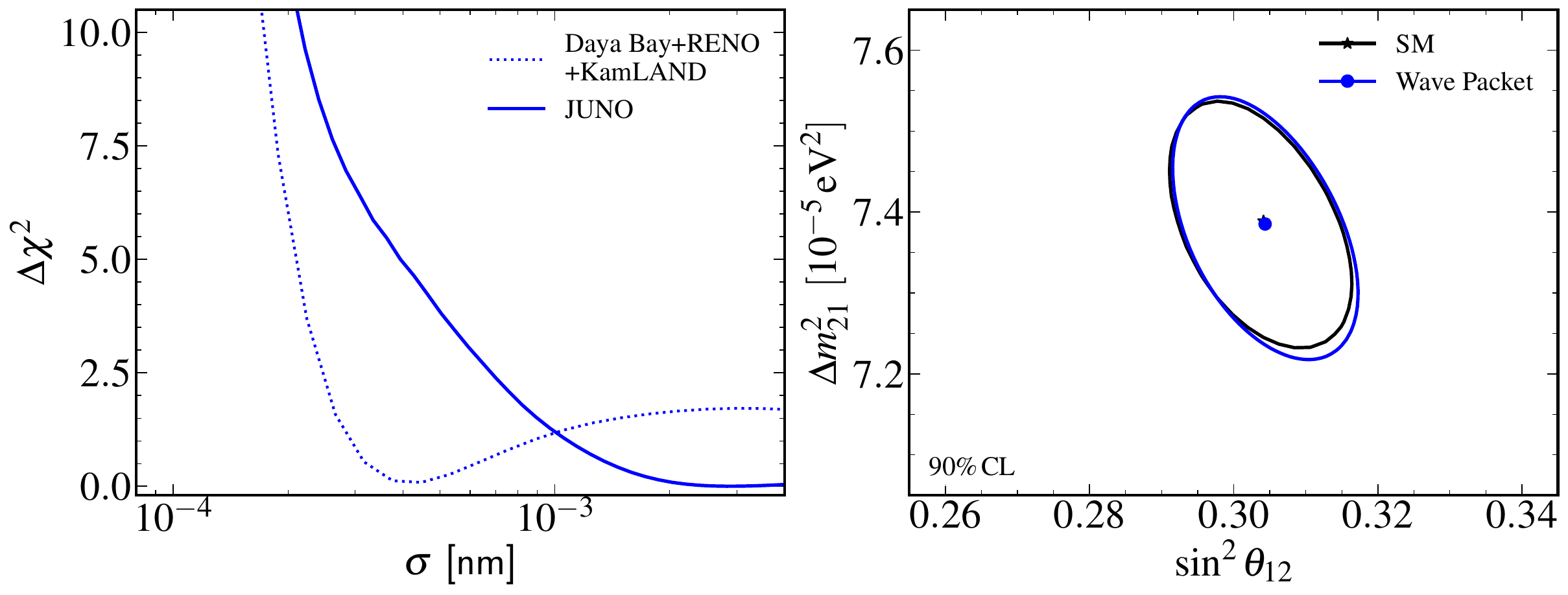}
    \caption{Left: $\Delta\chi^2$ profile for the wave packet size as obtained from the analysis of JUNO data (solid blue) in comparison with the combined bound (dashed blue) from previous reactor experiments~\cite{deGouvea:2024syg}. Right: Impact of the wave packet width on the determination of the standard solar parameters. Black contours correspond to the standard analysis, while the blue ones are obtained after marginalizing over $\sigma$.}
    \label{fig:sigma}
\end{figure}

We first discuss the potential of JUNO to bound the neutrino wave packet width. The result is shown in Fig.~\ref{fig:sigma}, where in the left panel we plot the bound that can be obtained from JUNO data, while in the right panel we focus on the measurement of the solar neutrino oscillation parameters.
{In the analysis we include an external constraint on $\sin^2\theta_{12}$ and $\Delta m_{21}^2$ from solar data, which are not affected by wave packet effects. This is justified, since solar neutrinos arrive as an incoherent superposition of mass states at the detectors on Earth, hence the measurement is robust under decoherence effects due to wave packet separation. In addition we include a constraint on $\Delta m_{31}^2$ from a combined analysis of long-baseline accelerator and atmospheric data~\cite{deSalas:2020pgw}. This can be done since the wave packets of the neutrinos relevant for these experiments are expected to be much larger than for reactor neutrinos. Their determination of $\Delta m_{31}^2$ is therefore not affected.  Since the determination of $\sin^2\theta_{13}$ is fairly robust in the combined analysis of reactor experiments~\cite{deGouvea:2021uvg}, we include a penalty on this quantity.We also verified that the bound obtained below is not affected even when leaving $\sin^2\theta_{13}$ free.
The resulting $\Delta\chi^2$ profile is shown in the left panel of Fig.~\ref{fig:sigma} (blue solid). Also shown is the combined bound (blue dotted) from the analysis of Daya Bay, RENO and KamLAND data for comparison~\cite{deGouvea:2021uvg,deGouvea:2024syg}.
As can be seen, with only 207 days of data taking, JUNO can place a stronger limit than all of the previous experiments combined. We find that JUNO does not see any minimum, contrary to the one at $\sigma\approx4\times10^{-4}$~nm that was obtained in Ref.~\cite{deGouvea:2024syg} for the other reactor experiments.}
The new bound on the wave packet width that can be obtained from the JUNO data at 90\% confidence level (CL) is

\begin{equation}
    \sigma >6.48\times10^{-4}~\textrm{nm}\,.
\end{equation}

We also discuss the effect of a finite wave packet width on the determination of the standard neutrino oscillation parameters. In the right panel of Fig.~\ref{fig:sigma} we show the allowed regions at 90\% CL for two degrees of freedom in the solar plane after minimizing over $\sigma$ (blue) in comparison with the standard analysis (black). {The analysis includes external constraints on $\sin^2\theta_{12}$ and $\Delta m_{21}^2$ from the solar data and on $\Delta m^2_{31}$ from long-baseline accelerators, atmospheric neutrino experiments and short-baseline reactors, as discussed before. The determinations of the solar parameters are very robust under this scenario.}

\subsection{Neutrino environmental decoherence}\label{sec:res-dec}

\begin{figure}[t!]
    \centering
    \includegraphics[width=1 \textwidth]{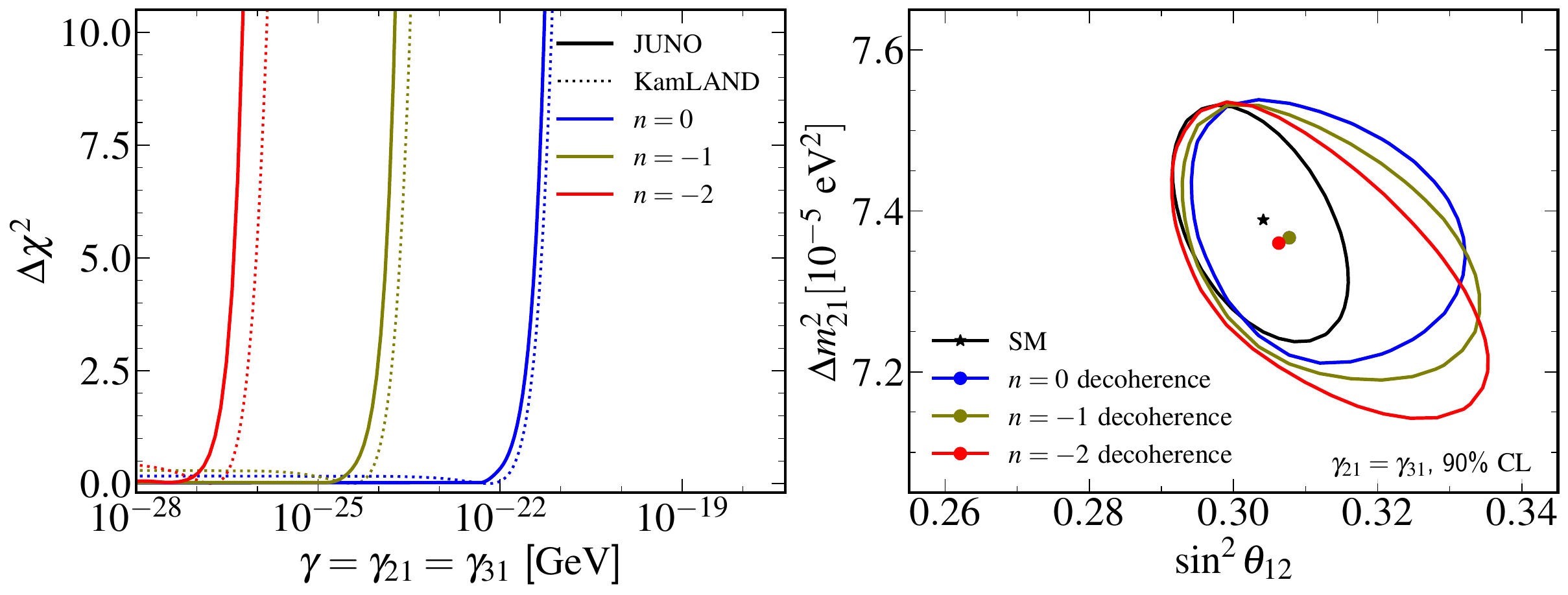}
\caption{Left: $\Delta\chi^2$ profiles for the decoherence parameters assuming $\gamma_{21} = \gamma_{31}$ and $\gamma_{32}=0$ from JUNO data {(solid lines)} in comparison with the bounds from KamLAND~\cite{DeRomeri:2023dht} {(dashed lines) for different values of $n$}. Right: The impact on the determination of the standard oscillation parameters.}
\label{fig:decoherence_g21g31}
\end{figure}

Next, we consider neutrino decoherence from interactions with some unknown environment. Note that the models that lead to Eq.~\eqref{eq:osc_damp} with the damping terms given in Eq.~\eqref{eq:D_lindblad}, sometimes called phase-perturbation models~\cite{Stuttard:2020qfv}, can not be distinguished from decoherence due to wave packet separation in the fully decohered regime. However, they lead to different behavior in the intermediate regime, where damping begins to become relevant, due to the different energy and baseline dependencies of the damping terms. 
The solar neutrino measurement of $\sin^2\theta_{12}$ {and $\Delta m_{21}^2$} is again robust under this scenario, {because the neutrinos which exit the Sun are already decohered. Solar experiments are sensitive to relaxation effects~\cite{deHolanda:2019tuf} present in state-selection models~\cite{Stuttard:2020qfv}. In this case one should handle the prior with care. We do not consider these effects and focus only on phase-perturbations and can safely include the solar penalty term in Eq.~\eqref{eq:chi2}. On the other, we do not include penalties on $\Delta m_{31}^2$ and $\sin^2\theta_{13}$ and minimize over them freely.}
{We first consider the case where $\gamma_{21} = \gamma_{31}$ and $\gamma_{32} = 0$. Note that, assuming $\gamma_{21} = \gamma_{32}$ and $\gamma_{31} = 0$ would lead to similar results and therefore it is not analyzed separately. Instead, taking $\gamma_{31} = \gamma_{32}$ and $\gamma_{21} = 0$ modifies only the fast oscillations, to which JUNO begins to become sensitive and this case will also be considered.}

The results of our analyses assuming $\gamma_{21} = \gamma_{31}$ and $\gamma_{32}=0$ are shown in Fig.~\ref{fig:decoherence_g21g31} for different choices of $n$. Since reactor neutrino energies are much smaller than $E_0 = 1$~GeV in Eq.~\eqref{eq:D_lindblad}, the weakest bound is obtained for $n=0$ (blue) and becomes stronger for decreasing $n$ (olive for $n=-1$ and red for $n=-2$). At 90\% confidence level, we find that

\begin{eqnarray}
    \gamma &<& 2.87 \times 10^{-22}~ \textrm{GeV}\,,\quad(n=0)\,,\\
    \gamma &<& 1.10\times 10^{-24}~ \textrm{GeV}\,,\quad(n=-1)\,,\\
    \gamma &<& 3.75 \times10^{-27}~ \textrm{GeV}\,,\quad(n=-2)\,.    
\end{eqnarray}
Despite the very limited exposure currently available, JUNO already reaches sensitivities slightly stronger than those obtained in Ref.~\cite{DeRomeri:2023dht} from the full KamLAND data set, which read $\gamma \lesssim 4\times10^{-22}$~GeV for $n=0$, $\gamma \lesssim 2\times10^{-24}$~GeV for $n=-1$ and $\gamma \lesssim 8\times10^{-27}$~GeV for $n=-2$.

In the right panel of Fig.~\ref{fig:decoherence_g21g31} we show the effect on the determination of the standard oscillation parameters. In the case of $n=0$ (same color code as in the left panel), the decoherence term is energy independent and we observe a correlation only with $\sin^2\theta_{12}$. For decreasing numbers of $n$, energy-dependence becomes more relevant and we observe also an effect on the determination of $\Delta m_{21}^2$. 

\begin{figure}[t!]
    \centering
    \includegraphics[width=1 \textwidth]{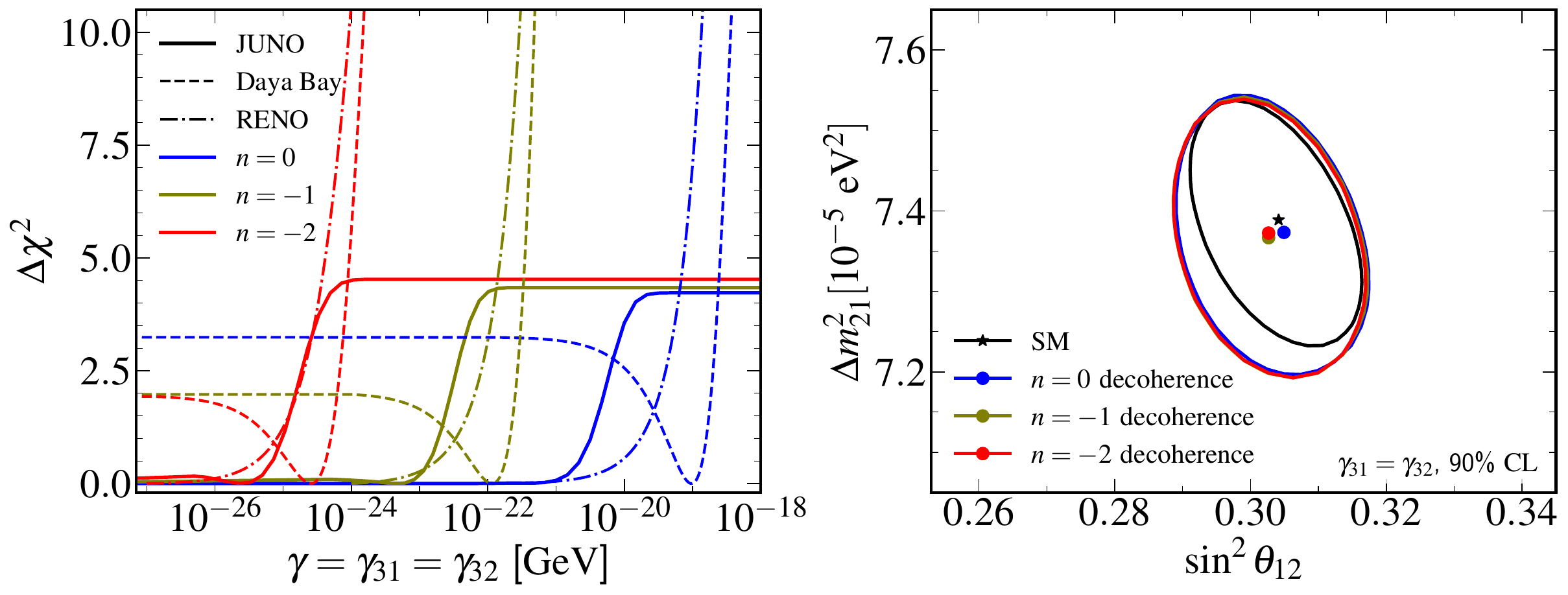}
\caption{Left: $\Delta\chi^2$ profiles for the decoherence parameters assuming $\gamma_{31} = \gamma_{32}$ and $\gamma_{21}=0$ from JUNO data {(solid lines)} in comparison with the updated bounds from Daya Bay {(dashed lines)} and RENO {(dashed-dotted lines)}. Right: The impact on the determination of the standard oscillation parameters.}
\label{fig:decoherence_g31g32}
\end{figure}

{For $\gamma := \gamma_{31} = \gamma_{32}$ and $\gamma_{21} = 0$ our results are presented in Fig.~\ref{fig:decoherence_g31g32}. The 90\% CL bounds are}

\begin{eqnarray}
    \gamma &<& 6.73 \times 10^{-21}~ \textrm{GeV}\,,\quad(n=0)\,,\\
    \gamma &<& 3.77\times 10^{-23}~ \textrm{GeV}\,,\quad(n=-1)\,,\\
    \gamma &<& 2.07 \times10^{-25}~ \textrm{GeV}\,,\quad(n=-2)\,.    
\end{eqnarray}
{As can be seen from the figure, the profiles rise but then form a plateau at $\Delta\chi^2\lesssim 5$. The reason is the still limited sensitivity to resolve the small wiggles in the spectrum produced by the fast oscillations. With more statistics and the expected energy resolution of 3\%, these bounds will behave in the same way as those shown in Fig.~\ref{fig:decoherence_g21g31}. 
We compare the bounds from JUNO with those obtained from the final data sets of the RENO~\cite{RENO:2025wbu} and Daya Bay~\cite{DayaBay:2022orm} experiments. These are shown as dashed-dotted and dashed lines in Fig.~\ref{fig:decoherence_g31g32} and update those of Ref.~\cite{DeRomeri:2023dht}, which relied on smaller data sets for both experiments. 
At 90\% CL JUNO provides already slightly stronger bounds than these experiments. However, unlike RENO and Daya Bay, current JUNO data is not capable to provide bounds at larger confidence levels. Note that our analysis of Daya Bay data exhibits a slight preference for decoherence effects. The same effect was already visible in Fig.~\ref{fig:sigma}. It is statistically not very significant and disfavored by RENO and JUNO.

In the right panel of Fig.~\ref{fig:decoherence_g31g32} we show the effect on the determination of the standard oscillation parameters, which does not differ for different values of $n$. The determination of both parameters is only mildly affected.}

\subsection{Invisible decay}

\begin{figure}[t!]
    \centering    \includegraphics[width=0.9
    \textwidth]{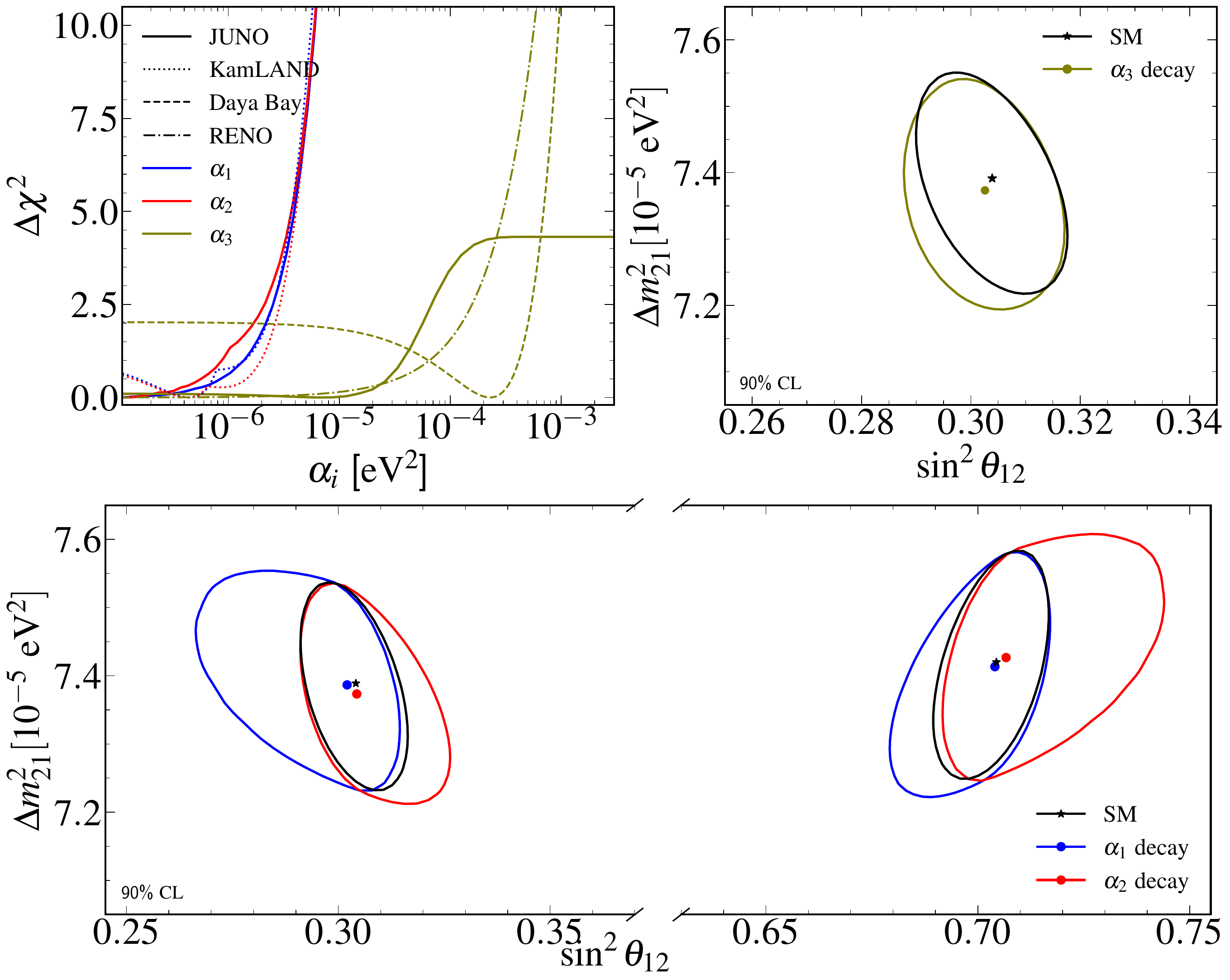}
    \caption{Upper left: $\Delta\chi^2$ profiles for the decay parameters $\alpha_i = m_i/\tau_i$ from the analysis of JUNO data {(solid lines) in comparison with the bounds of  RENO (dashed-dotted lines) and Daya Bay (dashed lines). Upper right: The impact of invisible decay of $\nu_3$ on the determination of the standard oscillation parameters. Lower panel: The impact of invisible decay of $\nu_1$ and $\nu_2$ on the determination of the standard oscillation parameters. In this case it is not possible to exclude the upper octant since external priors from solar data can not be used.}}
    \label{fig:decay}
\end{figure}

Finally, we discuss the bounds on neutrino invisible decay and its impact on the determination of the solar oscillation parameters. The results of our analyses are shown in Fig.~\ref{fig:decay}. {In this case, we cannot use a solar prior when analyzing $\alpha_1$ or $\alpha_2$, since the determination of solar parameters is affected due to the invisible decay of $\nu_1$ or $\nu_2$. Therefore we include it only in the analysis of $\alpha_3$.
On the other hand, the determination of $\sin^2\theta_{13}$ at short-baseline reactor experiments is always affected by invisible neutrino decay and we can therefore not include the penalty in any of the cases. Also $\Delta m_{31}^2$ is minimized over freely here.} In the upper left panel we present the bounds for $\alpha_1$ in blue, $\alpha_2$ in red, and $\alpha_3$ in olive. 
Note that the profiles are very similar for $\alpha_1$ and $\alpha_2$. The reason is that we allow for $\sin^2\theta_{12}$ to go to the second octant and the oscillation probability in Eq.~\eqref{eq:osc_damp} is approximately symmetric under the exchange of $\alpha_1\leftrightarrow\alpha_2$ and $\sin^2\theta_{12}\leftrightarrow\cos^2\theta_{12}$. {The sensitivity on $\alpha_3$ is still limited, as was the case of environmental decoherence with $\gamma_{31}=\gamma_{32}$, and it is expected to improve with more statistics. For this reason, even if the bound from JUNO is stronger than other reactor experiments at 90\% CL, it is not possible to obtain bounds at larger confidence levels.}
The bounds that can be placed on the decay parameters at 90\% CL are

\begin{eqnarray}
    \alpha_1 &<& 2.66\times10^{-6}~\textrm{eV}^2\,,\\
    \alpha_2 &<& 2.27\times10^{-6}~\textrm{eV}^2\,, \\
    \alpha_3 &<& 7.32\times10^{-5}~\textrm{eV}^2\,.
\end{eqnarray}
While they are of similar strength as bounds from other terrestrial oscillation probes~\cite{Ternes:2024qui,Gonzalez-Garcia:2008mgl}, the bounds from solar neutrino experiments on $\alpha_1$ and $\alpha_2$ are much stronger~\cite{Berryman:2014qha,SNO:2018pvg}. Nevertheless, reactor measurements provide an independent and complementary probe that relies on very different systematics and assumptions. {As shown in the lower panel of Fig.~\ref{fig:decay}, since for $\alpha_1$ and $\alpha_2$ it is not possible to use external constraints from solar data, the parameter space in the upper octant is still allowed. This does not happen for $\alpha_3$, where solar constraints are used, with the allowed regions shown in the upper right panel of the figure. While the measurement of solar parameters is fairly robust in the presence of $\alpha_3$, we find that the impact of $\alpha_1$ and $\alpha_2$ is of similar strength as in the decoherence scenarios with $n\neq0$.}

\section{Conclusion}
\label{sec:conc}

\begin{table}[t!]
\centering
\begin{subtable}{0.29\textwidth}
\centering
\begin{tabular}{|c|c|}
\hline
~~$\sigma$~[nm]~~ & ~~{$6.48\times10^{-4}$}~~ \\
\hline
$\alpha_1$~[eV$^2$] & ${2.66}\times 10^{-6}$ \\
$\alpha_2$~[eV$^2$] & ${2.27}\times10^{-6}$ \\
$\alpha_3$~[eV$^2$] & ${7.32}\times10^{-5}$ \\
\hline
\end{tabular}
\end{subtable}
\begin{subtable}{0.4\textwidth}
\centering
\begin{tabular}{|c|c|c|}
\hline
& $\gamma_{21}=\gamma_{31}$ [GeV] & $\gamma_{31}=\gamma_{32}$ [GeV]\\
\hline
$n=0$  & $2.87 \times 10^{-22}$ & $6.73 \times 10^{-21}$ \\
$n=-1$ & $1.10 \times 10^{-24}$ & $3.77 \times 10^{-23}$ \\
$n=-2$ & $3.75 \times 10^{-27}$ & $2.07 \times 10^{-25}$ \\
\hline
\end{tabular}
\end{subtable}
\caption{\label{tab:bounds} The bounds at 90\% CL ($\Delta\chi^2 = 2.71$) for the different scenarios considered in this paper.}
\label{tab:combined}
\end{table}

In this paper, we have investigated the sensitivity of the JUNO oscillation data to several scenarios that lead to a damping of neutrino oscillations, namely wave packet separation, environmental decoherence, and invisible neutrino decay. Although they arise from very different physical mechanisms, their effects can be studied within a unified damping framework, allowing a direct comparison of their phenomenological impact. 
We have used JUNO data to place bounds on the BSM parameters characterizing these models and studied their impact on the determination of the standard neutrino oscillation parameters $\sin^2\theta_{12}$ and $\Delta m_{21}^2$. The summary of the different bounds is presented in Tab.~\ref{tab:bounds}. 
These bounds are already competitive with other ones obtained in the literature as discussed in Sec.~\ref{sec:res}, and can be significantly improved in the future, in particular once JUNO {strengthens the determination of} the fast oscillations~\cite{deGouvea:2020hfl,DeRomeri:2023dht,Abrahao:2015rba,JUNO:2021ydg}. 

\begin{figure}[t!]
    \centering
    \includegraphics[width=1 \textwidth]{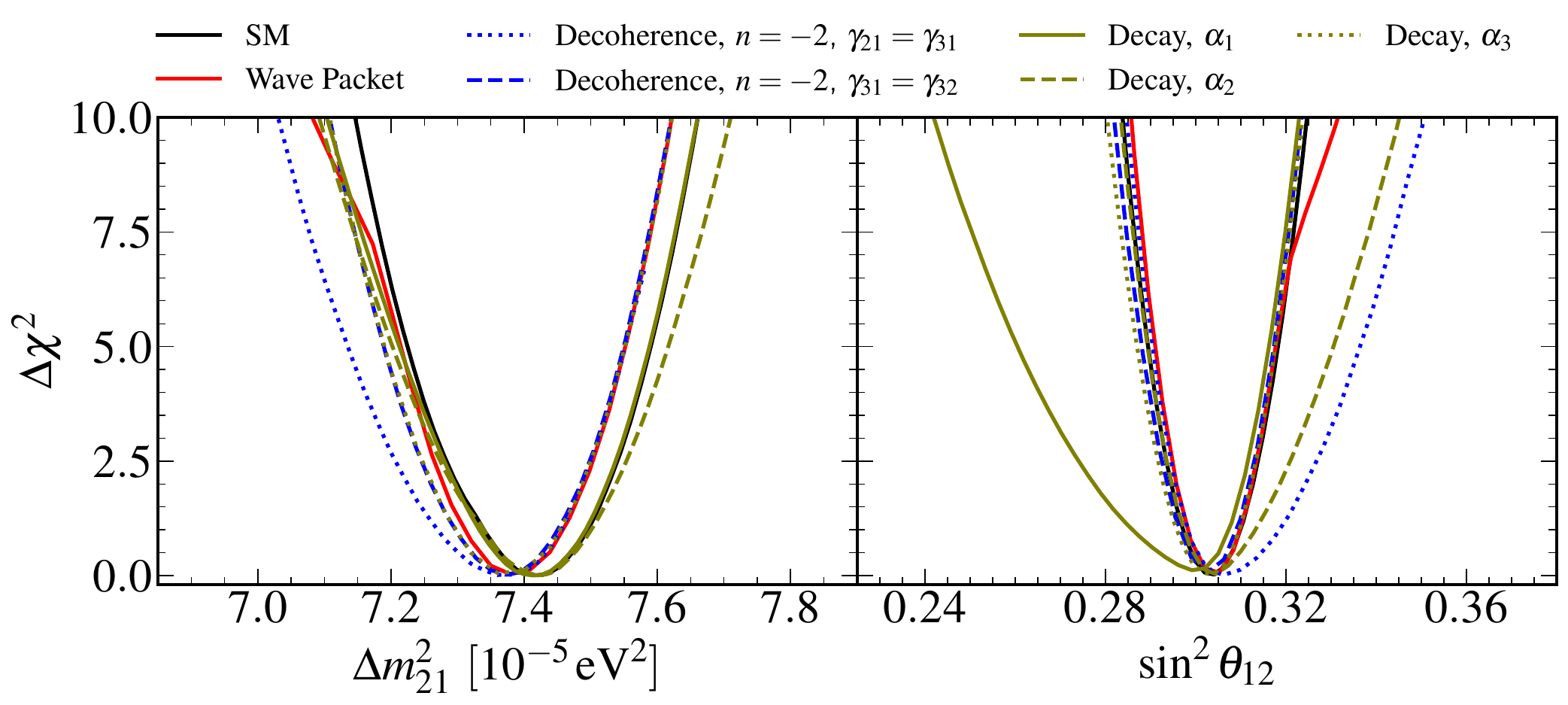}
    \caption{$\Delta\chi^2$ profiles for the standard oscillation parameters obtained from the different analyses performed in this paper.}
    \label{fig:chi2}
\end{figure}

Finally, in Fig.~\ref{fig:chi2} we present the $\Delta\chi^2$ profiles for the standard oscillation parameters $\Delta m_{21}^2$ (left) and $\sin^2\theta_{12}$ (right) of our JUNO analyses. We find that the determination of $\Delta m^2_{21}$ remains remarkably robust under all damping scenarios considered in this work. {The determination of $\sin^2\theta_{12}$ can be still considered robust, even if there is a non-negligible effect in the presence of invisible decay (for $\nu_1$ or $\nu_2$) or environmental decoherence (for $n=-2$ with $\gamma_{21} = \gamma_{31}$).
In any case,} as JUNO accumulates more statistics and particularly, after it {improves its ability to resolve} the atmospheric oscillation pattern, the sensitivity to damping signatures is expected to improve significantly, making it one of the most powerful probes of such effects.

\bibliographystyle{utphys}
\bibliography{bibliography}  

\end{document}